# Effect of Sr-for-Ba isovalent substitution on the local structure, hole distribution and magnetic irreversibility of Cu(Ba,Sr)$_2$YbCu$_2$O$_{6.95(2)}$

T. Nakane[a,#], K. Isawa[b], R. S. Liu[c], J. M. Chen[d], H. Yamauchi[a], and M. Karppinen[a,*]

[a]*Materials and Structures Laboratory, Tokyo Institute of Technology, Yokohama 226-8503, Japan*
[b]*R & D Center, Tohoku Electric Power Co., Inc., Sendai 981-0952, Japan*
[c]*Department of Chemistry, National Taiwan University, Taipei, Taiwan, R.O.C.*
[d]*National Synchrotron Radiation Research Center, Hsinchu, Taiwan, R.O.C.*

___

**Abstract**

The effect of Sr$^{II}$-for-Ba$^{II}$ isovalent substitution on the magnetic irreversibility field ($H_{irr}$) of Cu(Ba$_{1-y}$Sr$_y$)$_2$YbCu$_2$O$_{6.95(2)}$ (Cu-1212) sample series ($y = 0 \sim 0.4$) is studied to reveal guiding rules for tailoring the intrinsic $H_{irr}$ characteristics of high-$T_c$ superconductors. It has been assumed that substitution of the larger alkaline earth cation, Ba$^{II}$, by the smaller, Sr$^{II}$, might improve the $H_{irr}$ characteristics as a consequence of the decrease in the thickness of nonsuperconductive blocking block (BB). However, results of the present work show that Sr substitution rather depresses the $H_{irr}$ characteristics of the Cu-1212-phase superconductors even though the thickness of BB decreases. Both the amount of excess oxygen and the overall positive charge are confirmed to remain constant upon the Sr substitution by wet-chemical and XANES analyses, respectively. However, from neutron diffraction data analysis it is found that Sr substitution breaks the conductive CuO chains in BB by shifting part of the excess oxygen atoms from the characteristic *b*-axis lattice site to the *a*-axis site. This is believed to decrease the concentration of mobile holes in the BB, as supported by the results of TEP measurements. The lower $H_{irr}(T)$ lines of the Sr-substituted samples may thus be attributed to the lower concentration of mobile holes in BB.

___________________________________________________________________

[#]Present address: Superconducting Materials Center, National Institute for Materials Science, Ibaraki 305-0047, Japan
[*]Corresponding author: karppinen@msl.titech.ac.jp

## 1. Introduction

To realize most of the potential applications of high-$T_c$ superconductors (HTSCs), enhancement in the magnetic-field irreversibility ($H_{irr}$) characteristics of the existing HTSC materials is indispensable. Especially, HTSC materials with high $H_{irr}$ values at high temperatures ($\geq$ 77 K) are desired. The magnitude of $H_{irr}$ defines the "irreversibility-field line", $H_{irr}(T)$, when plotted against temperature ($T$). [Note that for the sake of comparison, $H_{irr}(T)$ lines are often plotted against a so-called reduced temperature, 1-$T/T_c$.] The $H_{irr}(T)$ lines of HTSCs are linked to both intrinsic and extrinsic flux pinning properties. Experimental data for the effect of columnar defects on $H_{irr}$ characteristics of HTSCs have shown that introducing effective extrinsic pinning centers alone does not allow us to reach the highest $H_{irr}(T)$ values in the high-temperature regieme [1]. It is thus essential for obtaining a high-$H_{irr}$-characteristics material that the intrinsic $H_{irr}$ property is improved. Considering the crystal structure of HTSCs [2], *i.e.* piling up of a superconductive block (SB) and a nonsuperconductive blocking block (BB) along the *c*-axis direction, the ways so-far suggested for enhancing the $H_{irr}$ characteristics are roughly classified into two categories: (*i*) increasing the overall hole-doping level [3-5] to enhance the thermodynamical stability of magnetic fluxons in SB, and (*ii*) enhancing the conductivity of BB [6-8] and/or decreasing the thickness of BB [8-11] to improve the correlation of fluxons (through BB) between two adjacent SBs. In general, the $H_{irr}$ characteristics of HTSCs have been discussed with respect to electromagnetic anisotropy ($\gamma^2$) such that lowering the anisotropy enhances the $H_{irr}$ characteristics [1,8,12]. The $\gamma^2$ value, typically estimated from resistivity measurements as $\gamma^2 = \rho_c / (\rho_a \rho_b)^{1/2}$, has been considered as a reasonable indicator of the intrinsic portion of the $H_{irr}$ characteristics. However, in reality there are no straightforward guiding rules for the control of the $\gamma^2$ value to tailor the $H_{irr}$ characteristics of HTSCs. Therefore demanded is a concrete recipé for enhancing the $H_{irr}$ characteristics of HTSCs from the stand-point of material design and synthesis.

Previously we have shown that the $H_{irr}$ characteristics depend not only on the total amount of doped holes but also on their distribution over the unit cell [2,13]. The latter may be altered when changing the former but also through "isovalent" cation substitution. That is, isovalent cation substitution is likely to affect the hole distribution through lattice contraction/expansion without changing the total hole-doping level [14,15]. Recently we succeeded in obtaining very high (intrinsic) $H_{irr}$ values for a CuBa$_2$(Yb$_{0.6}$Lu$_{0.4}$)Cu$_2$O$_{6.93}$ sample by controlling the hole distribution by means of isovalent smaller-for-larger rare-earth (*R*) substitution in the CuBa$_2$*R*Cu$_2$O$_{6+z}$ (Cu-1212) phase [16]. The key in this was that decreasing the size of *R* promotes the partial shift of holes from the CuO chains of the $CuO_z$ charge reservoir to the $CuO_2$ plane. As the holes in the Cu-1212 system with the larger $R$s are more concentrated in the $CuO_z$ charge reservoir rather than in the $CuO_2$ plane, such shift results in a more homogeneous distribution of holes, being advantageous in terms of enhancing the $H_{irr}$ characteristics.

To replace $Ba^{II}$ partially by $Sr^{II}$ is another way to apply the isovalent cation substitution scheme to HTSCs. It induces a contraction in BB, and might therefore be expected to improve the $H_{irr}$ characteristics. Quite importantly, for the Hg-based HTSCs replacing the larger Ba with the smaller Sr was proven to be an effective way to enhance the $H_{irr}$ values [9,17-19]. The same has been dreamt for the Cu-1212 phase, as this phase shows the highest $H_{irr}(T)$ values among all HTSCs and is thus closest to practical applications. Nevertheless, our recent works indicated that the $H_{irr}$ characteristics of the $CuBa_2RCu_2O_{6+z}$ (Cu-1212) and $Cu_2Ba_2RCu_2O_{7+z}$ (Cu-2212) phases are rather depressed by the Sr substitution [20]. In fact, the first observation of the decrease in the critical current ($J_c$) characteristics in Cu-1212 with Sr substitution is from 1991 by Parent and Moreau [21]. Here we report the results of our systematic study on the $Cu(Ba_{1-y}Sr_y)_2YbCu_2O_{6.95(2)}$ (Cu-1212) system, to reveal that the depression in the $H_{irr}$ characteristics due to the Sr substitution most likely originates from the decrease in the concentration of mobile holes within the BB. (Note that the $R$ = Yb system was selected rather than the prototype $R$ = Y one, as it possesses higher intrinsic $H_{irr}(T)$ values [16].) We reach this conclusion after precisely probing the changes caused by Sr substitution in the local atomic structures by neutron powder diffraction (NPD), in the block-specific hole concentrations by x-ray absorption near-edge structure (XANES) spectroscopy, and in the concentration of mobile holes by thermoelectric power (TEP) measurement.

## 2. Experimental

*Sample synthesis*: Samples of $Cu(Ba_{1-y}Sr_y)_2YbCu_2O_{6+z}$ ($y$ = 0, 0.1, 0.2, 0.3 and 0.4; $z \approx 0.95$) were synthesized from cation-stoichiometric mixtures of high-purity $BaCO_3$ (99.99 %), $SrCO_3$ (99.9 %), $Yb_2O_3$ (99.99 %) and CuO (99.99 %) powders. Calcination was carried out in air for 12 hours at an optimized temperature (900 ~ 940 ºC) that had been carefully determined for each composition through preliminary experiments. For obtaining high-quality single-phase Cu-1212 samples with the smallest $R$s, *i.e.*, with $R$ = Yb in the present work, a crucially important procedure during the calcination was to spread the powder as thin (~1 mm) as possible on an $Al_2O_3$ plate to guarantee highly homogeneous conditions in terms of both temperature and atmosphere during the synthesis procedure. After calcination, the powder was pelletized and sintered for 48 hours in air at an appropriate temperature (900 ~ 940 °C). The pellets of the Cu-1212 phase were then crushed and fully oxygenated by annealing in flowing $O_2$ gas at 350 °C for 48 hours. The oxygenated samples were characterized for the phase purity and the lattice parameters by x-ray diffraction (XRD; MAC Science: MXP18VAHF[22]; CuK$_\alpha$ radiation), and for the precise oxygen content by a $Cu^+/Cu^{2+}$ coulometric titration technique [2]. The precision of the coulometric titration method applied was better than ±0.01 with respect to the amount of excess oxygen, $z$ [13,16,19].

*Superconductivity characteristics*: Superconductivity properties were measured for the samples by a SQUID magnetometer (Quantum Design MPMS-XL). In this work, an "effective" $T_c^{(e)}$ value is considered rather than the onset temperature of the diamagnetic signal. The $T_c^{(e)}$ value is defined at the temperature where the extrapolated (dia)magnetic susceptibility ($\chi$) *vs.* $T$ curve crosses the $\chi = 0$ level, assuming that the paramagnetic contribution in the normal state is negligibly small. This value of $T_c^{(e)}$ is close to the "mean" $T_c$ value [22] that takes into account the distribution of $T_c$ value owing to subtle inhomogeneity of oxygen distribution. The $H_{irr}$ value was determined by the procedure introduced elsewhere [23]. In brief, the magnetic field ($H$) dependence of superconducting critical current $J_c(H)$ is calculated *via* Bean's model from the measured magnetization ($M$) *vs.* $H$ hysteresis loops, and then the $H_{irr}$ value is determined from the $\log(H)$ *vs.* $J_c(H)$ plot as the $H$ value at the limit of $J_c \to 0$, since $J_c(H_{irr}) = 0$.

*Neutron powder diffraction*: Neutron powder diffraction experiments were carried out at room temperature at R2 reactor in Studsvik, Sweden. A neutron powder diffractometer consisting of a double Cu (220) monochromator system ($l$ = 1.470 Å) and 35 $^3$He detectors was used by scanning in 0.08° steps over the 2θ range of 4.00 to 139.92°. Vanadinium sample holders with the diameter of 6 mm were utilized to measure the powder samples of 2.5 ~ 4.0 g. The neutron flux at the sample was *ca.* $10^6$ cm$^{-2}$ s$^{-1}$. For the structure refinements, a Rietveld program, FULLPROF, was used. The nomenclature of the atomic positions is given in Fig. 1. Applying orthorhombic space group, *Pmmm* (No. 47), refinements were carried out with respect to the atomic coordinates and isotropic temperature factors. Full occupancy was assumed for every site except for the O(1) and O(5) sites. The isotropic temperature factors for the O(1) and O(5) sites were fixed at 0.1 Å$^2$. In a previous study on Cu(Ba,Sr)$_2$YCu$_2$O$_{6+z}$ samples Licci *et al.* [24] used ($x$, 1/2, 0) rather than the conventional position of (0, 1/2, 0) for the O(1) site. This was also tested in the present work, but no improvements were seen in the reliability factors, $R_p$, $R_{wp}$ and $c^2$. Therefore, we used the conventional position at $x$ = 0 for O(1). As the last step of the refinement, the partial occupancy of Sr at the Ba site was allowed to vary from the nominal value. Since this neither resulted in essential deviations from the nominal values nor improved the reliability factors, the partial occupancies of Ba and Sr were fixed to the nominal values in the final refinements. The refined structural data were utilized to estimate the concentration of holes in the $CuO_2$ plane, $p(CuO_2)$, via bond-valence-sum (BVS) [25] calculations. The BVS estimation for $p(CuO_2)$, i.e., $p(CuO_2)_{BVS}$, was calculated from the BVS values of in-plane Cu and O atoms, $V_{Cu(2)}$ and $V_{O(2/3)}$, through: $p(CuO_2)_{BVS} = V_{Cu(2)} + V_{O(2)} + V_{O(3)} + 2$ [2,26]. The thicknesses of SB [$d$(SB)] and BB [$d$(BB)] were calculated using the refined position of Cu(2) as the border of the two blocks.

*X-ray absorption near-edge structure spectroscopy*: The O $K$-edge and Cu $L$-edge XANES measurements were performed in a fluorescence-yield mode on the 6-m High-Energy Spherical Grating Monochromator (HSGM) beam-line at National Synchrotron Radiation Research Center (NSRRC) in Hsinchu, Taiwan. The spectra were recorded from powder samples at room temperature using a micro-channel-plate (MCP) detector system consisting of a dual set of MCPs with an electrically isolated grid mounted in front of them. The photon energies were calibrated with an accuracy of 0.1 eV using the O $K$-edge absorption peak at 530.1 eV and the Cu $L$ white line at 931.2 eV of a CuO reference. The monochromator resolution was set to ~0.22 eV and ~0.45 eV at the O $K$ and Cu $L$ absorption edges, respectively. The spectral features were assigned and analyzed following Refs. 27 and 28. In the Cu $L_3$-edge area, the main peak is seen around 931 eV. This is due to divalent copper. The main peak has a shoulder at ~932.6 eV that reflects the amount of trivalent copper. In the analysis of the Cu $L_3$ spectral features, both the 931-eV peak and its shoulder were fitted with Gaussian functions. From the relative (integrated) intensity of the shoulder [$I(Cu^{III})$] against that of the main peak [$I(Cu^{II})$] an estimation for the average valence of copper [$V(Cu)_{XAS}$] was obtained as: $V(Cu)_{XAS} = 2 + I(Cu^{III}) / [I(Cu^{II}) + I(Cu^{III})]$. In the O $K$-edge spectra two pre-edge peaks are seen at ~527.5 and ~528.2 eV. These are due to hole states at the oxygen sites in BB and SB, respectively. Each spectrum was first normalized to have the same intensity for the main peak at ~537.5 eV, and then the pre-edge peaks were fitted with Gaussian functions. The intensities obtained for the peaks at ~527.5 eV and at ~528.2 eV were supposed to reflect the relative amounts of holes in the BB [$p(BB)_{XAS}$] and in the two $CuO_2$ planes, [$2p(CuO_2)_{XAS}$], respectively. In order to make the relative numbers obtained from the O $K$-edge XANES analyses more illustrative, they were normalized such that the sum, $p(BB)_{XAS} + 2p(CuO_2)_{XAS}$, for the $y = 0$ sample equals to the amount of holes *per* formula unit as calculated from the precise oxygen content of the sample, *i.e.* $z = 0.94$. The average Cu-O layer hole concentration [$p(Cu-O)_{XAS}$] that provides us with a convenient measure of the average doping level of the phase was calculated as: $p(Cu-O)_{XAS} = [2p(CuO_2)_{XAS} + p(BB)_{XAS}] / 3$.

*Thermoelectric power measurement*: For the TEP measurements a two-channel thermoelectric measurement system was utilized in the temperature range of 30 ~ 305 K, employing two pairs of copper-constantan thermocouples for the measurement of the temperature gradient ($\Delta T$) over the sample. The magnitude of $\Delta T$ was kept at < 0.5 K over the gap (~2.5 mm). Absolute values of the Seebeck coefficient ($S$) were calculated from the relative measurement data by subtracting the contribution from the copper reference wires. The uncertainty in the measured values was estimated at ±0.3 $\mu$V/K.

## 3. Results and Discussion

From the x-ray diffraction data, the synthesized Cu(Ba$_{1-y}$Sr$_y$)$_2$YbCu$_2$O$_{6+z}$ samples were of single phase up to $y = 0.3$. The sample with $y = 0.4$ was found to contain traces of secondary phases, Yb$_2$BaCuO$_5$ and BaCuO$_2$, and was therefore not characterized by NPD. It should be noted that full Sr-for-Ba substitution is possible in Cu-1212 through high-pressure synthesis, but the high-pressure synthesized CuSr$_2$YCu$_2$O$_{6+z}$ compound is not anymore totally isostructural with CuBa$_2$YCu$_2$O$_{6+z}$ [29]. The NPD data confirmed the conclusion made based on the XRD data, that the samples with $0 \leq y \leq 0.3$ were all free from impurity phases. In Fig. 2, shown is the fitted NPD pattern for the $y = 0.3$ sample. The values of the oxygen content as independently determined from NPD data and from wet-chemical analysis, the $T_c$ values and the structural parameters refined from the NPD data for the samples are summarized in Table 1. The values refined for the oxygen content from the NPD data are highly consistent with those revealed from the wet-chemical analysis. The both measurements verify that the oxygen content remains constant at $z = 0.95(2)$ within the sample series. Thus it is concluded that the total amount of positive charge remains constant upon Sr substitution. The same is seen from the results of Cu $L_3$-edge and O $K$-edge XANES data. In Figs. 3(a) and 3(b), the average valence of copper, $V$(Cu)$_{XAS}$, as estimated from the Cu $L_3$-edge spectrum and the average hole concentration of the three Cu-O layers (*i.e.* one CuO$_z$ layer and two CuO$_2$ planes), $p$(Cu-O)$_{XAS}$, as estimated from the O $K$-edge spectrum are plotted against the Sr-substitution level, $y$. Both plots confirm that the overall hole-doping level remains constant within the error limits of the XANES analyses. In other words, Sr substitution works in a manner expected for ideal isovalent substitution.

Figure 4 displays the evolution of thicknesses of the two blocks, SB and BB, with increasing Sr content, $y$. The former increases whereas the latter decreases as $y$ increases. Since the BB contraction is larger in magnitude than the expansion of SB, the $c$-axis length shrinks upon Sr substitution as expected. The contraction of BB is apparently due to the smaller size of $Sr^{II}$ as compared with that of $Ba^{II}$. On the other hand, the increase in the thickness of SB with increasing Sr content is likely to be caused by increased hole-doping level of the $CuO_2$ plane. To illustrate this, we plot in Fig. 5 the values of $p(CuO_2)_{BVS}$ and $p(CuO_2)_{XAS}$ against the Sr content. Despite the slightly scattered data for $p(CuO_2)_{XAS}$, it is obvious that these two independent parameters that both reflect the $CuO_2$-plane hole concentration increase with increasing $y$. This means that Sr substitution promotes a shift of holes from the $CuO_z$ charge reservoir to the $CuO_2$ planes. A recent NQR (nuclear quadrupole resonance) study concluded the same for the $Cu(Ba,Sr)_2YCu_2O_{6+z}$ system [30]. In line with these results for the Cu-1212 phase, very similar consideration based on BVS calculation and NQR data has shown that Sr substitution results in a BB-to-SB hole shift in the $Cu_2(Ba,Sr)_2YCu_2O_8$ system of the Cu-2212 phase as well [31]. For the present samples, with increasing Sr content and thus increasing $p(CuO_2)$ the value of $T_c$ decreases (Table 1). This may well be attributed to the increasing degree of "overdoping" with increasing substitution level of Sr, though one should keep in mind that the value of $T_c$ is likely to be affected by other factors too, such as the buckling of the $CuO_2$ plane [32]. Here the buckling angle slightly increases with $y$ (from 7.92° to 8.32°).

From the observed shrinkage of the BB and shift of holes from the BB to the $CuO_2$ plane, one would expect enhanced $H_{irr}$ characteristics upon the Sr-for-Ba substitution in $Cu(Ba,Sr)_2YbCu_2O_{6+z}$ (*cf.* the discussion in Introduction). What we see, however, is that the $H_{irr}$ characteristics are not enhanced but rather depressed by Sr substitution. In Fig. 6 shown are the $H_{irr}$ values in terms of reduced temperature, $1-T/T_c$, for all the five $Cu(Ba_{1-y}Sr_y)_2YbCu_2O_{6.95(2)}$ samples. The most plausible explanation for the reduced $H_{irr}$ characteristics upon Sr substitution in Cu-1212 is based on the fact observed from the NPD data. That is, even though the total amount of excess oxygen, $z$, remains constant we see a gradual shift of the oxygen occupancy from the O(1) site to the O(5) site in the $CuO_z$ charge reservoir as Sr substitution proceeds (Table 1). This is illustrated in Fig. 7 in which the quantity, $\Delta z$, that represents the change in the oxygen occupancy is plotted for the two sites, O(1) and O(5), against $y$. For the $Cu(Ba_{1-y}Sr_y)_2YCu_2O_{6+z}$ system, the very first NPD study using a sample with $y = 0.5$ ($z$ not given) concluded full ordering of the oxygen vacancies [33], but later Licci *et al.* [24] revealed partial shift of the charge-reservoir oxygen from the *b*-axis site to the *a*-axis site due to Sr substitution. As a consequence of the O(1)-to-O(5) shift, the Cu(1)-O(1) chains get broken and also the Cu(1) atoms in these chains become inequivalent as some of them are directly coordinated not only to O(1) but also to O(5). Both the consequences are likely to depress the concentration of "mobile" holes and thus the electrical conductivity of the $CuO_z$ charge reservoir. Here it should also be noted that the Cu(1)-O(1) chains in Cu-1212 are responsible for conductivity along the *c*-axis direction [34,35].

Thermoelectric power measurement has proven to be a powerful tool to reveal the metallicity of (fully-oxygenated) CuO chains in Cu-1212 and related copper oxides [36-38]. As a pronounced example, the compound $Cu_2Ba_2PrCu_2O_8$ (Cu-2212) possesses "perfect" CuO (double) chains, for which the metallic conductivity was verified by TEP measurements [38]. Therefore we anticipate that the change in the concentration of mobile holes in our Cu-1212 samples due to Sr substitution might be visible from the TEP data. In Fig. 8, we show the temperature dependence of the Seebeck coefficient for the three $Cu(Ba_{1-y}Sr_y)_2YbCu_2O_{6.95(2)}$ samples at $y = 0.1$, 0.2 and 0.3. The overall behavior of $S(T)$ is similar to that previously reported for fully-oxygenated samples of the $CuBa_2YbCu_2O_{6+z}$ phase [39,40], *i.e.* $S$ is quite small in magnitude and falls into negative values above $T_c$. These characteristics are typical for overdoped HTSCs [41]. From the reported works on the Cu-1212 and Cu-2212 phases, we have learned that the CuO chain contribution yields positive $\Delta S/\Delta T$, whereas that of the $CuO_2$ plane is negative [36-42]. In other words, for fully-oxygenated Cu-1212 samples with highly conducting chains seen is an upward tail that develops in the high-temperature region above approximately 150 K to the direction of increasing temperature. For the present samples in the high-temperature region, both $\Delta S/\Delta T$ and the magnitude of $S$ (clearly though not strongly) decrease with increasing $y$, indicating that the concentration of mobile holes in the $CuO_z$-chain charge reservoir that contribute to TEP is decreased. For example, we may look at the temperature range from 150 K to 250 K and estimate $\Delta S/\Delta T$ from Fig. 8 at 0.0195, 0.0125 and 0.0090 $\mu V/K^2$ for $y = 0.1$, 0.2 and 0.3, respectively. This behavior is qualitatively similar to that already seen for the $Cu(Ba,Sr)_2YCu_2O_{6+z}$ system with Sr substitution [42]. Thus we conclude that our TEP data are in line with our expectation based on the NPD data for the charge-reservoir oxygen arrangement, that the mobile-hole concentration in BB decreases in our $Cu(Ba_{1-y}Sr_y)_2YbCu_2O_{6.95(2)}$ sample series as the Sr-substitution level increases. We furthermore suggest that this is the principal reason for the suppression of the $H_{irr}$ characteristics in this system. The present work has thus demonstrated that the $H_{irr}$ characteristics of HTSCs are not primarily controlled by the lattice dimension, but rather by changes in the layer-specific concentration(s) of (mobile) holes.

## 4. Conclusions

A series of Cu(Ba,Sr)$_2$YbCu$_2$O$_{6+z}$ samples of the Cu-1212 phase was prepared and characterized in respect to the effect of Sr substitution. The oxygen content of the samples was fixed to the constant value of $z = 0.95(2)$, such that also unchanged remained the overall "chemical" doping level as determined by the sample stoichiometry and probed by XANES spectroscopy. Despite the lattice contraction and the shrinkage of BB, both evidenced to occur from the refined NPD data as a consequence of Sr substitution, the $H_{irr}$ characteristics were found depressed. We therefore concluded that the thickness of BB itself is not of primary importance in terms of determining the $H_{irr}$ characteristics of a HTSC phase at a fixed (chemical) doping level. Moreover we showed (consistently from both the NPD and XANES data) that the charge redistribution that is likely to accompany the Sr$^{II}$-for-Ba$^{II}$ substitution (or any other isovalent substitution) did not decrease the CuO$_2$-plane hole concentration for our Cu(Ba,Sr)$_2$YbCu$_2$O$_{6+z}$ samples but rather worked to the opposite direction. (In the former case only decreased $H_{irr}$ values would be expected.) A plausible explanation for the depressed $H_{irr}(T)$ lines was obtained from the NPD data: the excess oxygen in the charge reservoir was found to reorganize in such a way that the concentration of perfect CuO chains decreased, presumably leading to a decrease in the concentration of mobile holes in BB. Here the TEP data was recognized to point be in line with. The reason for the lower $H_{irr}(T)$ lines of the Sr-substituted Cu-1212 samples may thus be the lower concentration of mobile holes in the charge reservoir.

**Acknowledgments**

This work was supported by the EC Large Scale Facility Project (Ref. No. 629) and also by a Grand-in-Aid for Scientific Research (Contract No. 11305002) from the Ministry of Education, Science and Culture of Japan. Prof. S. Eriksson and Dr. H. Rundlöf (Studsvik Neutron Research Laboratory, Uppsala University) are thanked for the neutron powder diffraction measurements. We also acknowledge Prof. T. Matsushita and Dr. T. Motohashi for helpful discussions.

# References

1. Y. Nakayama, T. Motohashi, K. Otzschi, J. Shimoyama, K. Kitazawa, K. Kishio M. Konczykowski, N. Chikumoto, Phys. Rev. B 62 (2000) 1452.
2. M. Karppinen, H. Yamauchi, Mater. Sci. Eng. R 26 (1999) 51.
3. K. Kishio, J. Shimoyama, T. Kimura, Y. Kotaka, K. Kitazawa, K. Yamafuji, Q. Li, M. Suenaga, Physica C 235-240 (1994) 2775.
4. K. Fujinami, H. Suematsu, M. Karppinen, H. Yamauchi, Physica C 307 (1998) 202.
5. M. Karppinen, H. Yamauchi, T. Nakane, M. Kotiranta, Physica C 338 (2000) 18.
6. J.L. Tallon, G.V.M. Williams, C. Bernhard, D.M. Pooke, M.P. Staines, J.D. Johnson, R.H. Meinhold, Phys. Rev. B 53 (1996) R11972.
7. K. Kishio, J. Shimoyama, A. Yoshikawa, K. Kitazawa, O. Chmaissem, J.D. Jorgensen, J. Low Temp. Phys. 105 (1996) 1359.
8. J.L. Tallon, G.V.M. Williams, J.W. Loram, Physica C 338 (2000) 9.
9. J. Shimoyama, S. Hahakura, R. Kobayashi, K. Kitazawa, K. Yamafuji, K. Kishio, Physica C 235-240 (1994) 2795.
10. V. Hardy, A. Maignan, C. Martin, F. Warmont, J. Provost, Phys. Rev. B 56 (1997) 130.
11. M.P. Raphael, M.E. Reeves, E.F. Skelton, C. Kendziora, Phys. Rev. Lett. 84 (2000) 1587.
12. K. Kishio, In: *Coherence in High Temperature Superconductors*, G. Deutscher, A. Revcolevschi (Eds.), World Scientific, Singapore 1996, pp. 212-225.
13. T. Nakane, K. Fujinami, M. Karppinen, H.Yamauchi, Supercond. Sci. Technol. 12 (1999) 242.
14. M. Karppinen, H. Yamauchi, Int. J. Inorg. Mater. 2 (2000) 589.
15. M. Karppinen, N. Kiryakov, Y. Yasukawa, T. Nakane, H. Yamauchi, Physica C 382 (2002) 66.
16. Y. Yasukawa, T. Nakane, H. Yamauchi, M. Karppinen, Appl. Phys. Lett. 78 (2001) 2917.
17. S. Lee, N.P. Kiryakov, D.A. Emelyanov, M.S. Kuznetsov, Yu.D. Tretyakov, V.V. Petrykin, M. Kakihana, H. Yamauchi, Yi Zhuo, M.-S. Kim, S.-I. Lee, Phys Rev. B 305 (1998) 57.
18. N. Kiryakov, S. Lee, M. Karppinen, H. Yamauchi, K. Yamawaki, S. Sasaki, Physica C 357-360 (2001) 350.
19. A.J. Batista-Leyva, M.T.D. Orlando, L. Rivero, R. Cobas, E. Altshuler, Physica C 383 (2003) 365.
20. T. Nakane, M. Karppinen, H. Yamauchi, Physica C 357-360 (2001) 226.
21. L. Parent, C. Moreau, J. Mater. Sci. 26 (1991) 5873 .

22. T. Matsushita, E.S. Otabe, T. Nakane, M. Karppinen, H. Yamauchi, Physica C 322 (1999) 100.
23. H. Yamauchi H. Yamauchi, M. Karppinen, K. Fujinami, T. Ito, H. Suematsu, H. Sakata, K. Matsuura, K. Isawa, Supercond. Sci. Technol. 11 (1998) 1006.
24. F. Licci, A. Gauzzi, M. Marezio, G.P. Radaelli, R. Masini, C. Chaillout-Bougerol, Phys. Rev. B 58 (1998) 15208.
25. I.D. Brown, D. Altermatt, Acta Cryst. B 41 (1985) 244.
26. M. Karppinen, H. Yamauchi, K. Fujinami, T. Nakane, K. Peitola, H. Rundlöf, R. Tellgren, Phys. Rev. B 60 (1999) 4378.
27. N. Nücker, E. Pellegrin, P. Schweiss, J. Fink, S.L. Molodtsov, C.T. Simmons, G. Kaindl, W. Frentrup, A. Erb, G. Müller-Vogt, Phys. Rev. B 51 (1995) 8529.
28. M. Karppinen, H. Yamauchi, T. Nakane, K. Fujinami, K. Lehmus, P. Nachimuthu, R.S. Liu, J.M. Chen, J. Solid State Chem. 166 (2002) 229.
29. O.I. Lebedev, G. Van Tendeloo, F. Licci, E. Gilioli, A. Gauzzi, A. Prodi, M. Marezio, Phys. Rev. B 66 (2002) 132510.
30. X.N. Ying, B.Q. Li, Y.H. Liu, Y.N. Huang, Y.N. Wang, Phys. Rev. B 66 (2002) 12506.
31. J. Karpinski, S.M. Kazakov, M. Angst, A. Mironov, M. Mali, J. Roos, Phys. Rev. B 64 (2001) 94518.
32. M. Guillaume, P. Allenspach, W. Henggeler, J. Mesot, B. Roessli, U. Staub, P. Fischer, A. Furrer, V. Trounov, J. Phys.: Condens. Matter 6 (1994) 7963.
33. B.W. Veal, W.K. Kwok, A. Umezawa, G.W. Grabtree, J.D. Jorgensen, J.W. Downey, L.J. Nowicki, A.W. Mitchell, A.P. Paulikas, C.H. Sowers, Appl. Phys. Lett. 51 (1987) 279.
34. J.L. Tallon, C. Bernhard, U. Binninger, A. Hofer, G.V.M. Williams, E.J. Ansaldo, J.I. Budnick, Ch. Niedermayer, Phys. Rev. Lett. 74 (1995) 1008.
35. N.E. Hussey, H. Takagi, Y. Iye S. Tajima, A.I. Rykov, K. Yoshida, Phys. Rev. B 61 (2000) 6475.
36. J.-S. Zhou, J.P. Zhou, J.B. Goodenough, J.T. McDevitt, Phys. Rev. B 51 (1995) 3250.
37. C. Bernhard, J.L. Tallon, Phys. Rev. B 54 (1996) 10201.
38. I. Terasaki, N. Seiji, S. Adachi, H. Yamauchi, Phys. Rev. B 54 (1996) 11993.
39. W.N. Kang, M.Y. Choi, Phys. Rev. B 42 (1990) 2573.
40. K. Matsuura, T. Wada, Y. Yaegashi, S. Tajima, H. Yamauchi, Phys. Rev. B 46 (1992) 11923.
41. S.D. Obertelli, J.R. Cooper, J.L. Tallon, Phys. Rev. B 46 (1992) 14928.
42. J.G. Lin, C.W. Chang, C.Y. Huang, C.Y. Chang, R.S. Liu, Physica C 336 (2000) 249.

**Table 1.** Oxygen contents, $T_c$ values and NPD refinement results for the $Cu(Ba_{1-y}Sr_y)_2YbCu_2O_{6+z}$ samples with $y = 0, 0.1, 0.2$ and $0.3$

| $y$ | | 0.0 | 0.1 | 0.2 | 0.3 |
|---|---|---|---|---|---|
| $n_{O(1)} + n_{O(5)}$ | titration | 0.92(1) | 0.93(1) | 0.96(1) | 0.95(1) |
| $n_{O(1)} + n_{O(5)}$ | refinement | 0.96(2) | 0.97(2) | 0.97(2) | 0.94(2) |
| $T_c$ [K] | | 88.4 | 84.6 | 80.8 | 77.9 |
| $a$ [in Å] | | 3.8037(2) | 3.8008(2) | 3.7940(2) | 3.7868(2) |
| $b$ [in Å] | | 3.8712(3) | 3.8680(3) | 3.8621(9) | 3.8554(3) |
| $c$ [in Å] | | 11.657(9) | 11.634(9) | 11.615(9) | 11.5904(9) |
| Yb | $B$ [in Å$^2$] | 0.25(5) | 0.33(5) | 0.19(4) | 0.25(5) |
| | $z$ | 0.1847(4) | 0.1843(4) | 0.1840(3) | 0.1842(4) |
| Ba/Sr | $n$ | 1.0/0.0 | 0.90/0.10 | 0.80/0.20 | 0.70/0.30 |
| | $B$ [in Å$^2$] | 0.24(9) | 0.24(9) | 0.48(8) | 0.59(10) |
| Cu(1) | $B$ [in Å$^2$] | 0.23(7) | 0.31(7) | 0.22(7) | 0.41(8) |
| Cu(2) | $z$ | 0.3568(3) | 0.3566(3) | 0.3558(2) | 0.3555(3) |
| | $B$ [in Å$^2$] | 0.08(6) | 0.07(6) | 0.06(5) | 0.12(6) |
| O(1) | $n$ | 0.959(8) | 0.948(8) | 0.939(8) | 0.910(8) |
| O(2) | $z$ | 0.3802(4) | 0.3795(4) | 0.3796(4) | 0.3784(4) |
| | $B$ [in Å$^2$] | 0.26(8) | 0.38(9) | 0.43(8) | 0.51(9) |
| O(3) | $z$ | 0.3809(5) | 0.3805(5) | 0.3795(4) | 0.3800(5) |
| | $B$ [in Å$^2$] | 0.52(8) | 0.51(9) | 0.60(8) | 0.63(9) |
| O(4) | $z$ | 0.1602(4) | 0.1602(4) | 0.1606(4) | 0.1610(4) |
| | $B$ [in Å$^2$] | 0.44(9) | 0.75(10) | 0.83(9) | 1.01(11) |
| O(5) | $n$ | 0.002(8) | 0.018(8) | 0.027(8) | 0.029(8) |
| $R_p$ [%] | | 8.09 | 9.50 | 8.64 | 9.82 |
| $R_{wp}$ [%] | | 9.69 | 10.6 | 9.63 | 11.1 |
| $c^2$ | | 3.39 | 3.09 | 4.60 | 3.80 |

*Note: Space group P4/mmm (No. 47): Yb at (0.5, 0.5, 0.5), Ba/Sr at (0.5, 0.5, z), Cu(1) at (0, 0, 0), Cu(2) at (0, 0, z), O(1) at (0, 0.5, 0), O(2) at (0.5, 0, z)), O(3) at (0, 0.5, z), O(4) at (0, 0, z) and O(5) at (0.5, 0, 0),.*

**FIGURE CAPTIONS**

**Fig. 1.** Crystal structure of $Cu(Ba,Sr)_2YbCu_2O_{6+z}$ and notation of the atoms.

**Fig. 2.** Fitting of the NPD pattern for the $Cu(Ba_{1-y}Sr_y)_2YbCu_2O_{6+z}$ sample with $y = 0.3$. The residual curve between the observed and calculated intensities is shown in the bottom.

**Fig. 3.** Changes (a) in average Cu valence, $V(Cu)_{XAS}$, as estimated from Cu $L_3$-edge XANES data, and (b) in average Cu-O layer hole concentration, $p(Cu\text{-}O)_{XAS}$, as estimated from O $K$-edge data, plotted against the Sr-substitution level, $y$, for $Cu(Ba_{1-y}Sr_y)_2YbCu_2O_{6.95(2)}$.

**Fig. 4.** Thicknesses of SB, $d(SB)$ (filled circles), and BB, $d(BB)$ (open circles), plotted against the Sr-substitution level, $y$, for $Cu(Ba_{1-y}Sr_y)_2YbCu_2O_{6.95(2)}$.

**Fig. 5.** $CuO_2$-plane hole concentration as estimated from O $K$-edge data, $p(CuO_2)_{XAS}$ (filled circles), and as calculated by BVS method from NPD data, $p(CuO_2)_{BVS}$ (open circles), plotted against the Sr-substitution level, $y$, for $Cu(Ba_{1-y}Sr_y)_2YbCu_2O_{6.95(2)}$.

**Fig. 6.** $H_{irr}$ characteristics of the $Cu(Ba_{1-y}Sr_y)_2YbCu_2O_{6.95(2)}$ ($0 \leq y \leq 0.4$) samples.

**Fig. 7.** Changes in the oxygen occupancy at the O(1) site (filled circles) and the O(5) site (open circles) in the $CuO_z$ charge reservoir for the $Cu(Ba_{1-y}Sr_y)_2YbCu_2O_{6.95(2)}$ samples plotted against the Sr-substitution level, $y$.

**Fig. 8.** Temperature dependence of Seebeck coefficient for the $Cu(Ba_{1-y}Sr_y)_2YbCu_2O_{6.95(2)}$ ($y = 0.1, 0.2, 0.3$) samples.

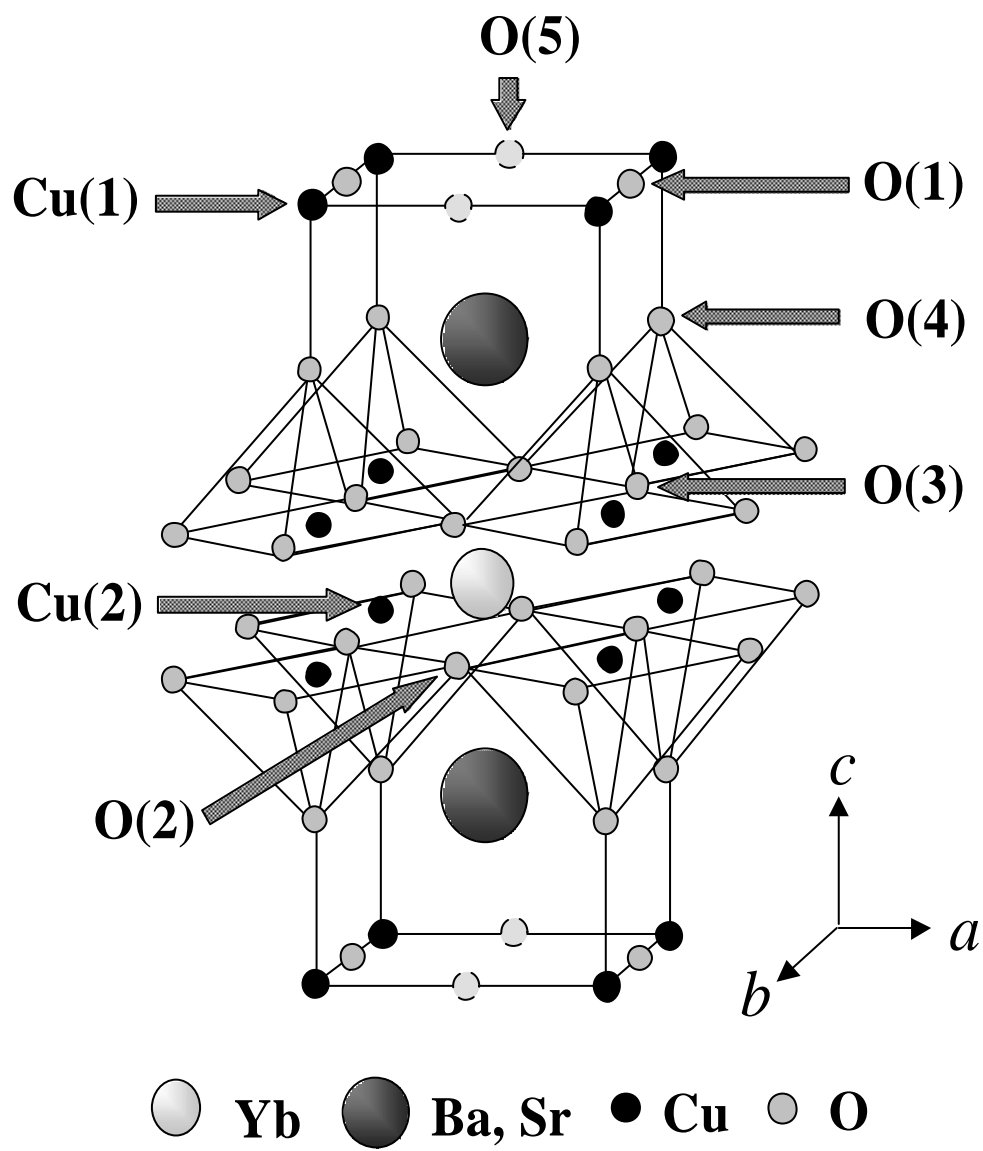

Nakane *et al*: Fig. 1.

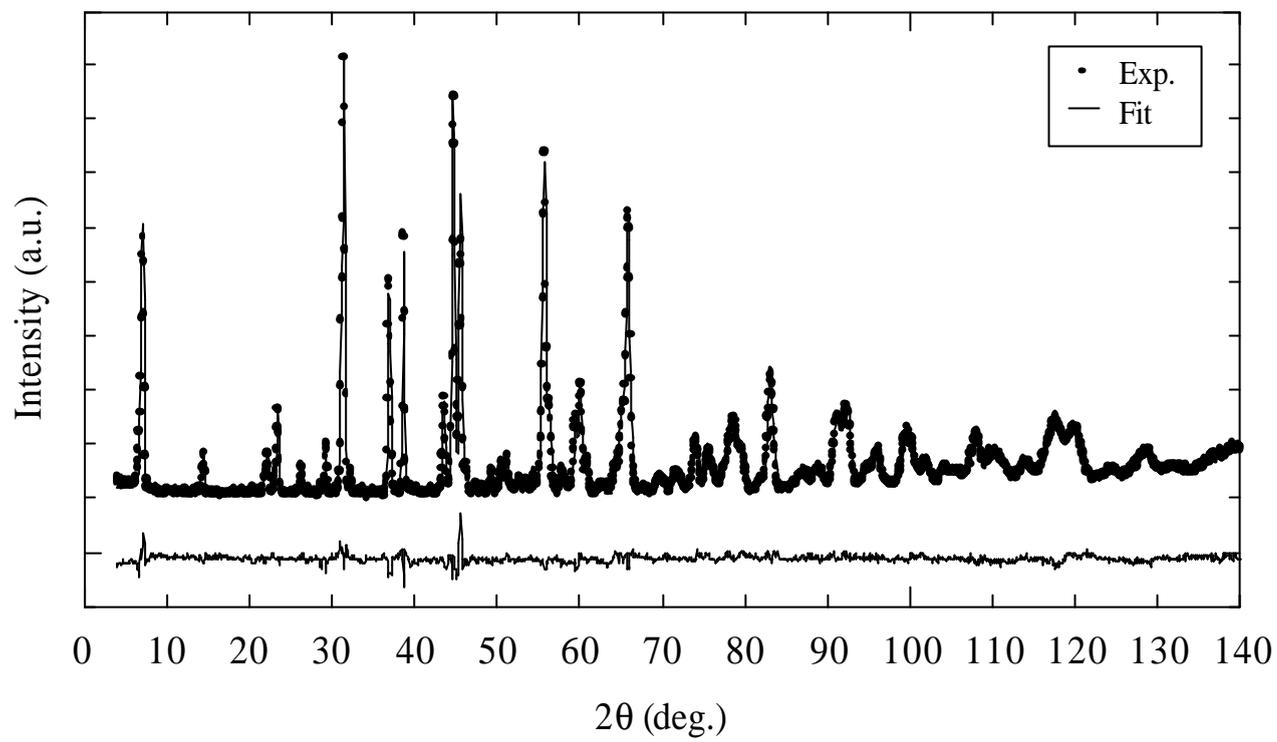

Nakane *et al*: Fig. 2.

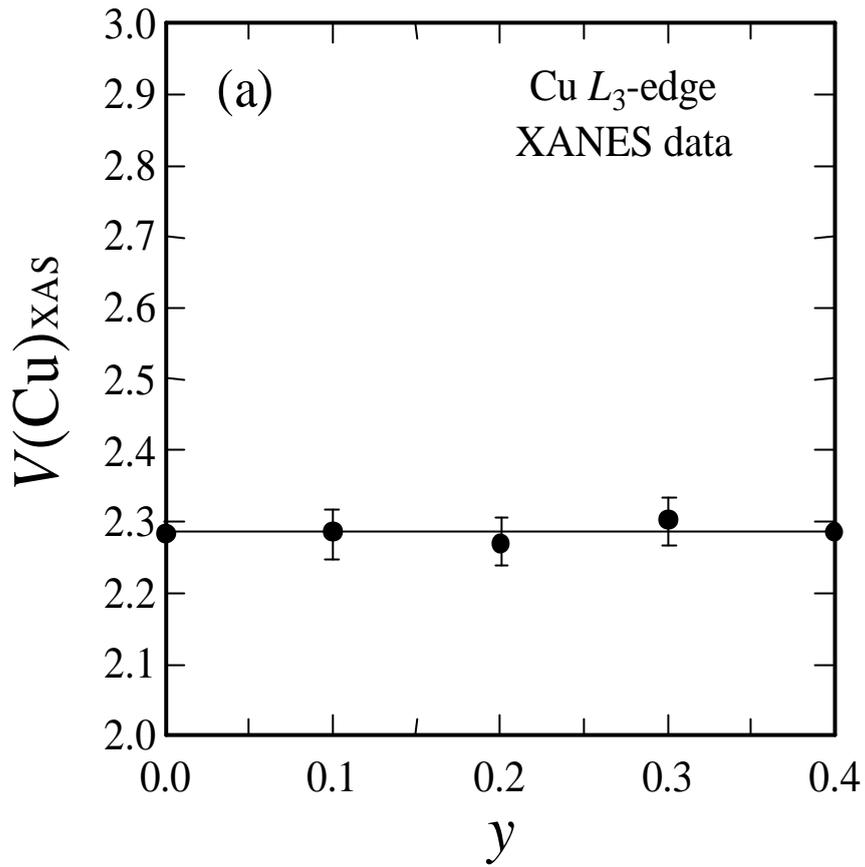
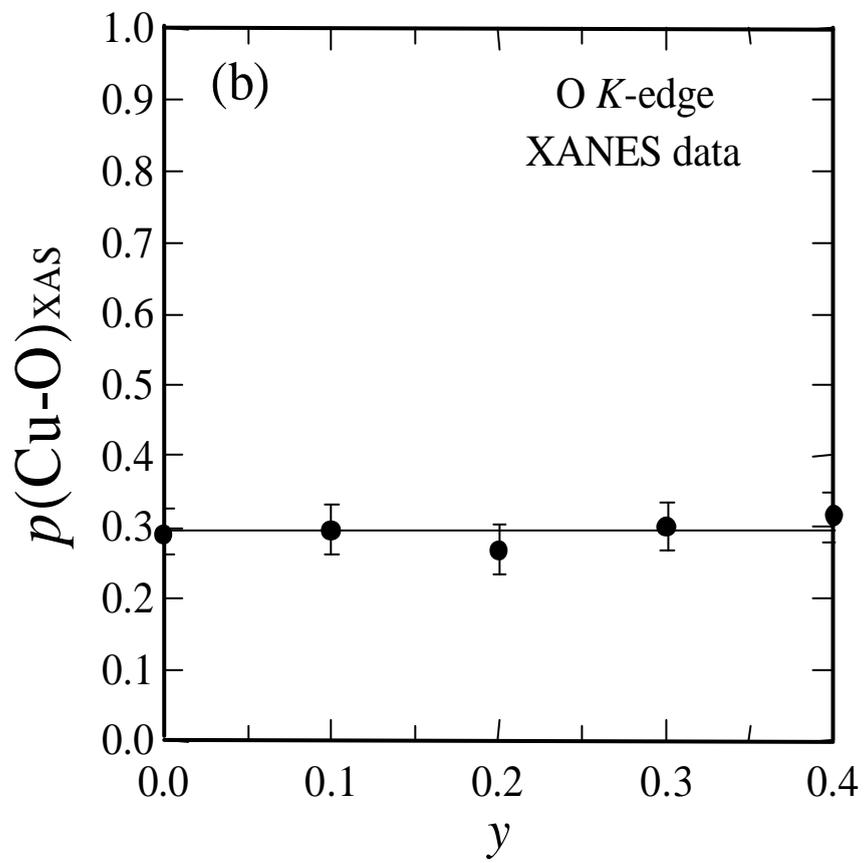

Nakane *et al*: Fig. 3.

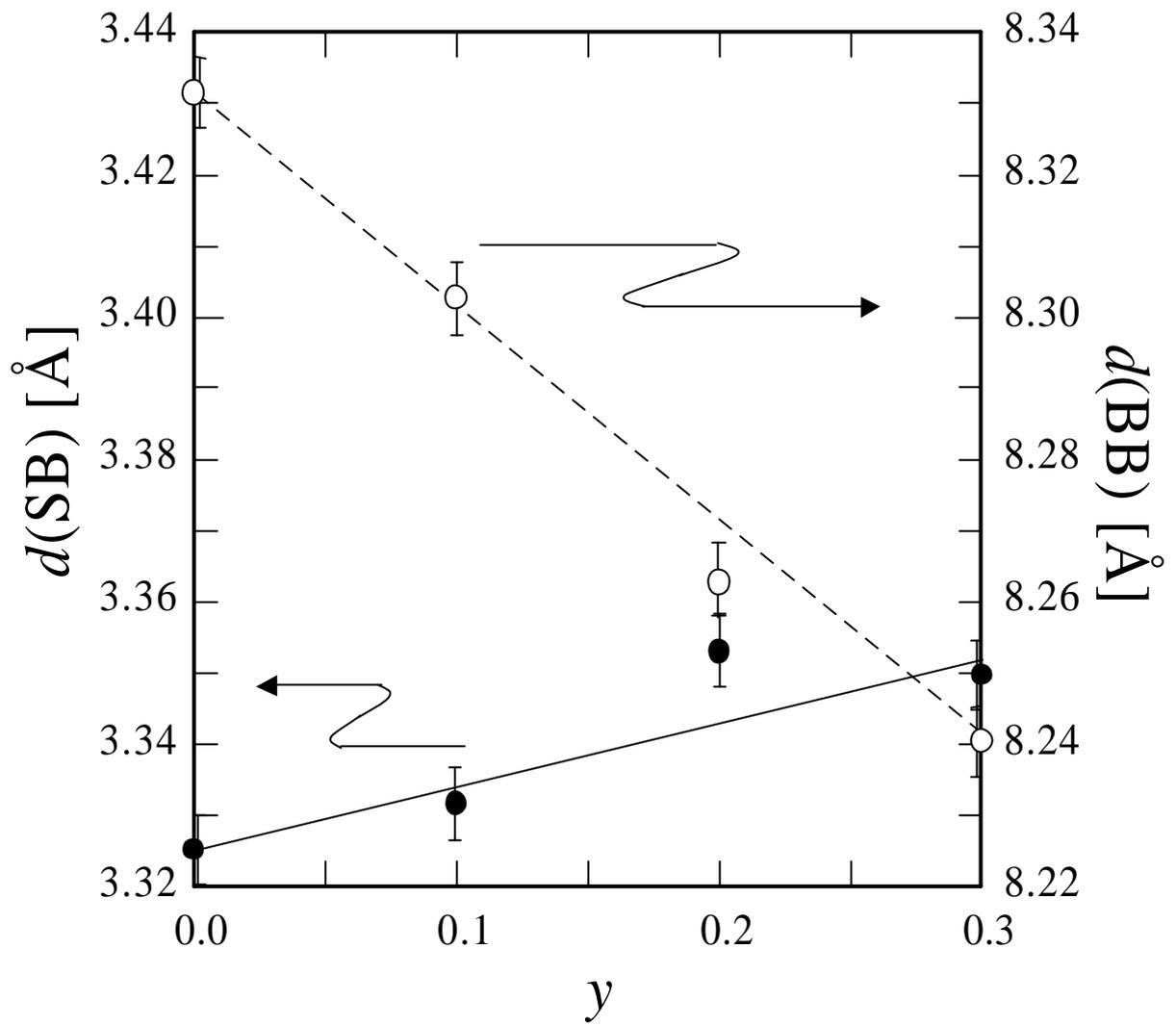

Nakane *et al*: Fig. 4.

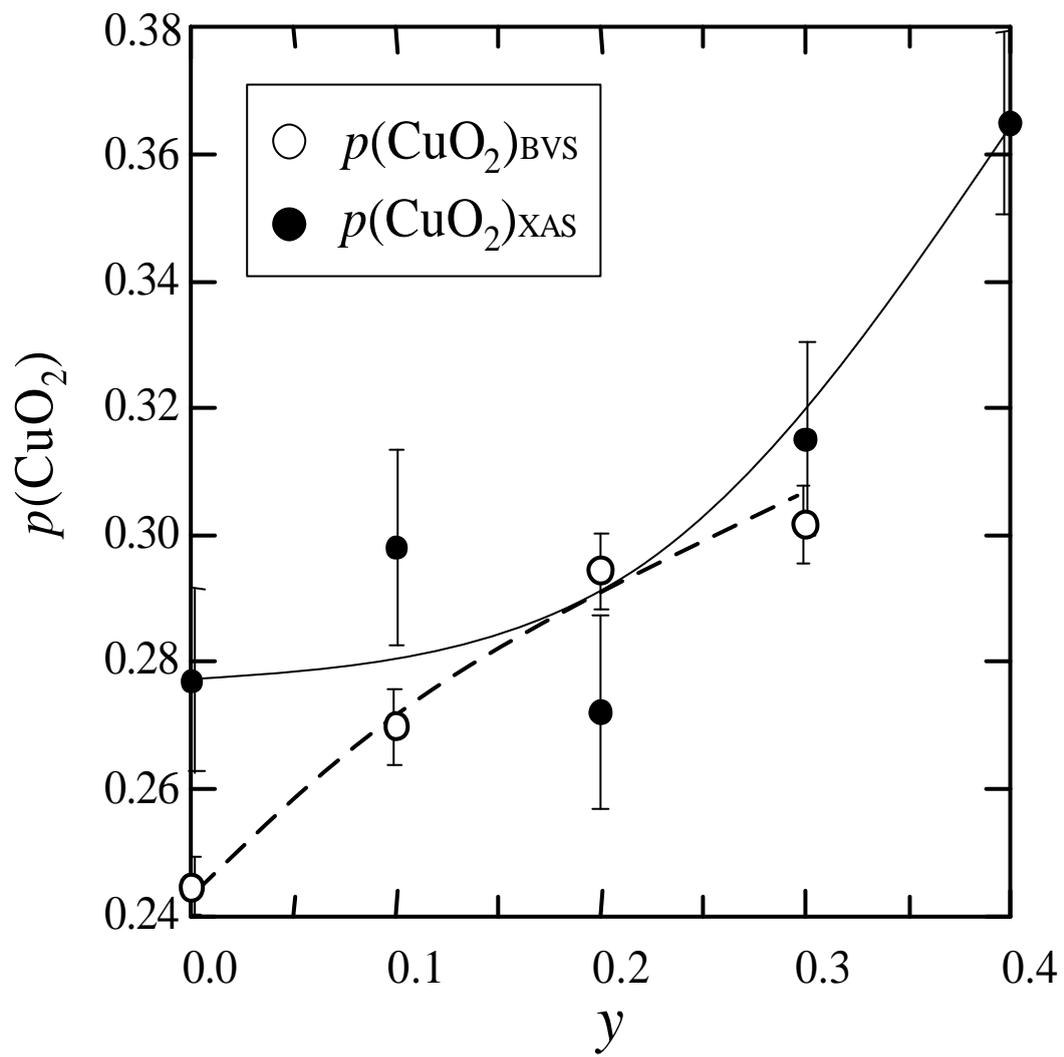

Nakane *et al*: Fig. 5.

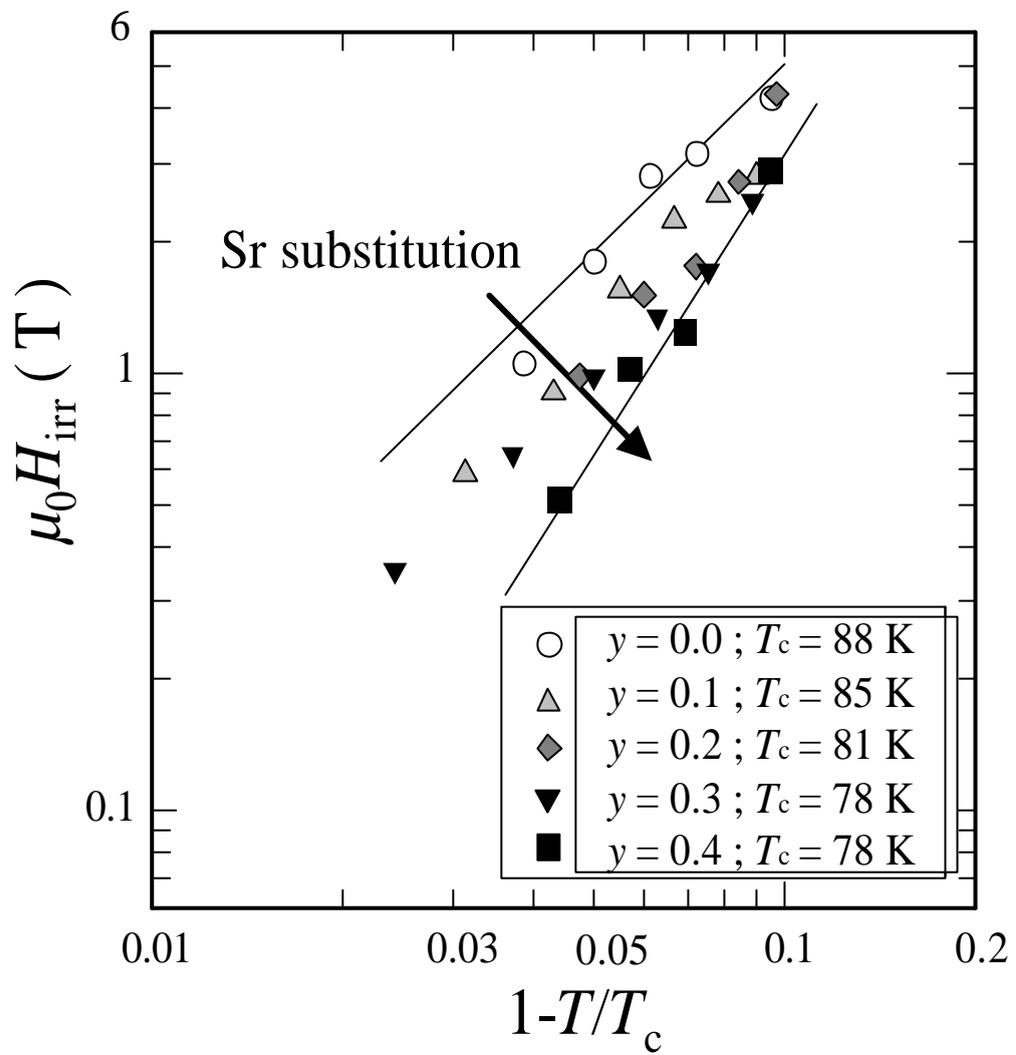

Nakane *et al*: Fig. 6.

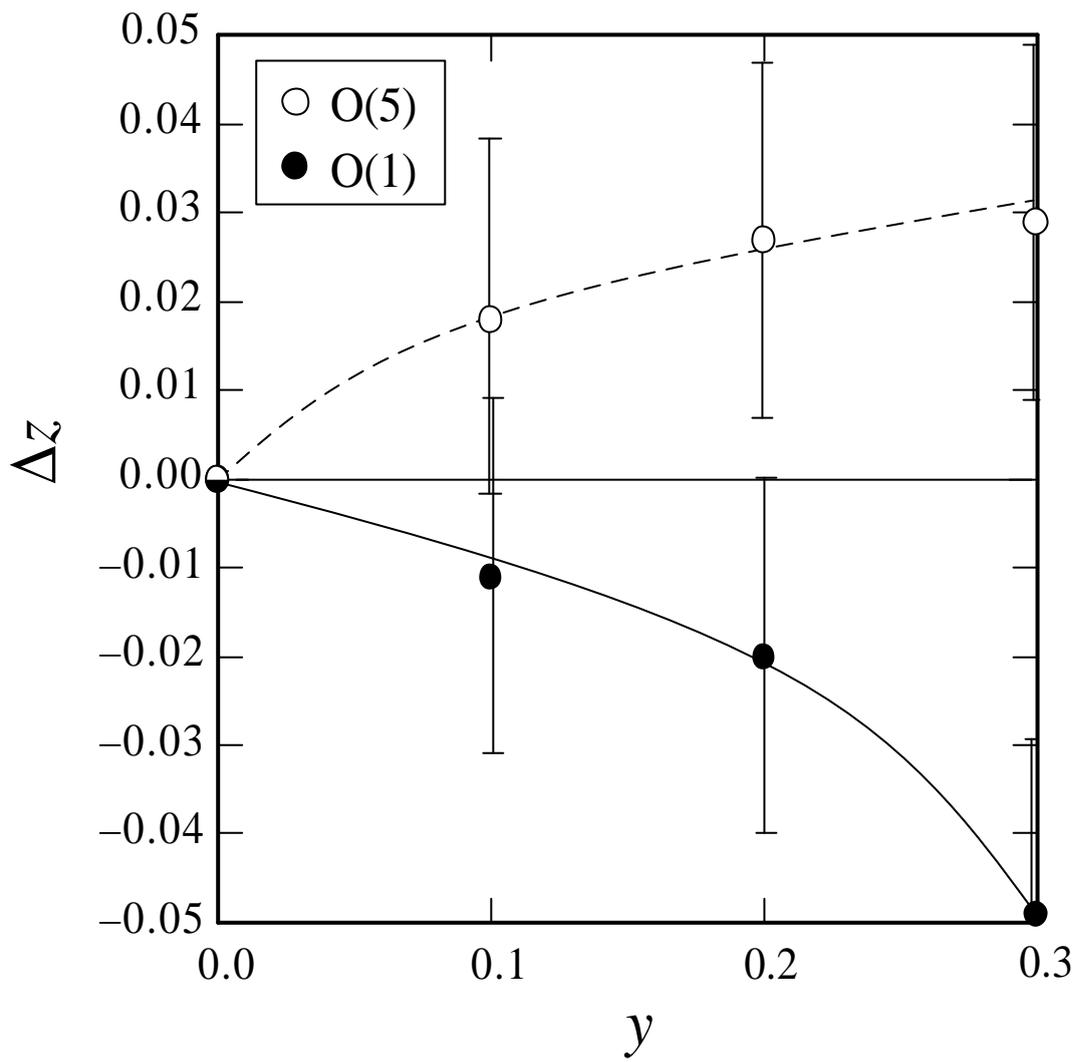

Nakane *et al*: Fig. 7.

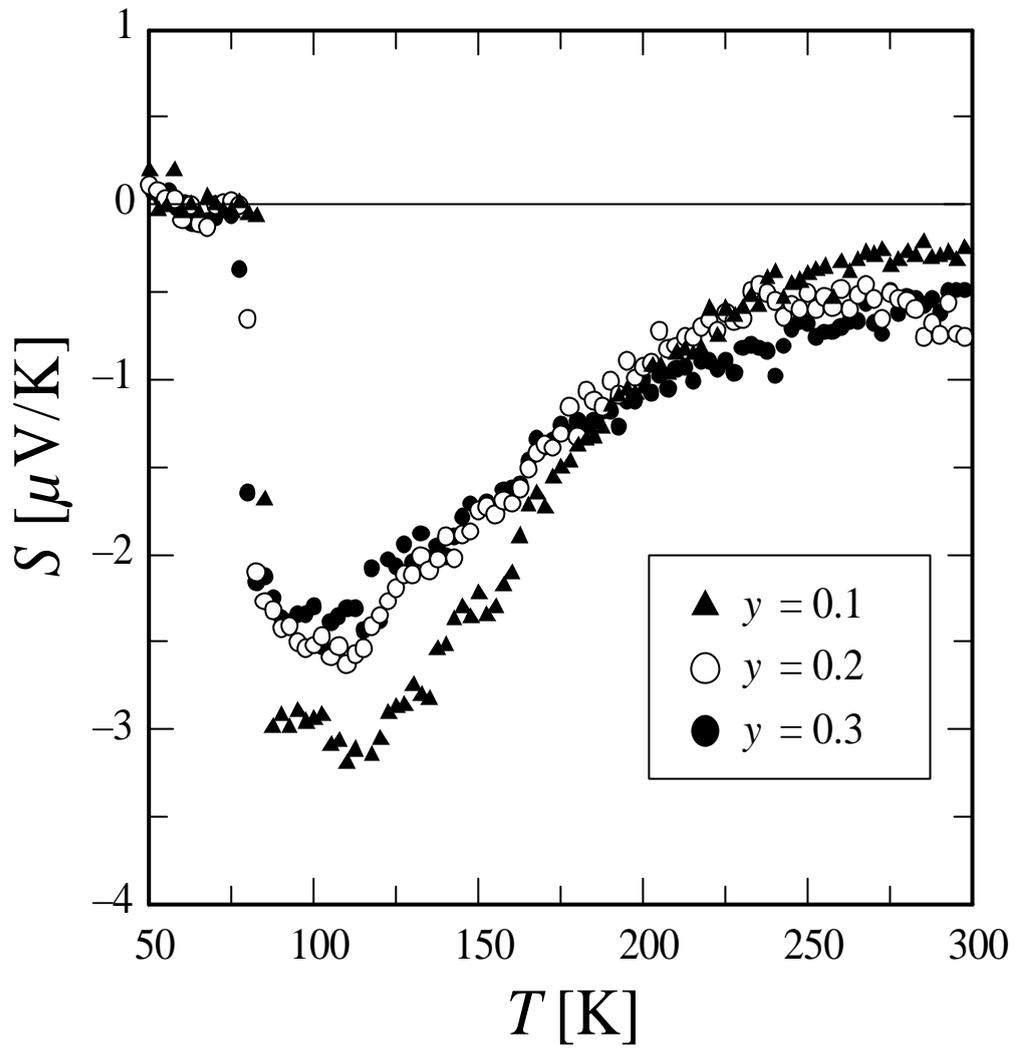

Nakane *et al*: Fig. 8.